\documentstyle[prl,aps,floats]{revtex}
\def\ep {\epsilon}

\def\e2 {\epsilon-\epsilon_k}
\def\be {\begin{equation}}
\def\ee {\end{equation}}
\def\bea {\begin{eqnarray}}
\def\eea {\end{eqnarray}}
\def\om {\omega}

\begin{document}

\hsize\textwidth\columnwidth\hsize\csname@twocolumnfalse\endcsname

\draft

\title{On the Normal State of the High Temperature Superconductors}

\bigskip
\author{George Kastrinakis}
\address{Department of Physics, University of Illinois at Urbana-Champaign,
1110 West Green St.,Urbana, IL 61801 }
\date{September 25, 1996}

\maketitle
\begin{abstract}
Based on a Fermi liquid model, we present several results on the normal
state of the optimally doped and overdoped cuprate superconductors.
Our main result is an analytic demonstration, backed by
self-consistent numerical calculations, of the linear in temperature
resistivity and linear in 1/(energy) optical conductivity, 
provided the interacting Fermi liquid has strong peaks in its density
of states (van-Hove singularities in 2 dimensions) near the chemical potential.
Moreover, we find that the interactions tend to pin these strong peaks
close to the chemical potential. This fact compares favorably with experiment
on a variety of cuprates. Finally, we show that the above scenario
yields naturally the low energy dependence of the experimentally determined
susceptibility, without reference to spin waves.
\end{abstract}
 

\vspace{1cm}

\centerline{\bf On the scattering rate of the cuprates}

\vspace{1cm}
	
	We consider a 2-dimensional fermionic model with a tight-binding dispersion
$\ep_k$
(with nearest neighbor hopping $t$, and next nearest neighbors of the
desired order), chemical potential $\mu$,
and some bare coupling constants $\{U_i\}$ (where the indices may stand
for the usual Hubbard $U$, a constant Coulomb coupling etc.) \cite{eg}.
This is a reasonable starting point for the description of the quasi-particles
in the planes of the optimally doped and overdoped cuprates.

	We write a diagrammatic Baym-Kadanoff self-consistent approximation
for this system, which can be solved numerically \cite{eg}.
The main characteristic of this approach of interest here, is that
the self-energy $\Sigma$ is given as the convolution
of the {\em dressed} Green's function $G$ times an effective potential $V$.
The latter arises as a combination of the couplings $\{U_i\}$ and the
susceptibility $\chi_o(q,\om_m)=-T \sum_{k,\ep_n}
G(k+q,\ep_n+\om_m) G(k,\ep_n)$ ($\ep_n, \om_m$ being Matsubara energies).

	The numerical solution of the many-body system yields always
an interesting result for the self energy at finite temperature. 
$Im \Sigma(k,\ep)$ turns out 
to be essentially linear in energy in an interval $\ep_1<\ep<\ep_2$. 
However it does have the correct parabolic Fermi liquid bevahior 
for $\ep \rightarrow 0$.
The parabolic region shrinks with decreasing filling $n$. Furthermore,
the energy interval of linear behavior expands as the 
van-Hove singularities ($\nabla \ep_k = 0$) at the points 
$q_o =(\pm \pi,0), (0,\pm \pi)$
become stronger (there are no other van-Hove singularities for the
dispersion we used).

	We give an analytic derivation of this result.
	It can easily be shown \cite{agd} that $Im \Sigma(k,\ep)$ 
is given by the following formula at finite temperature :
\be
Im \Sigma^R(k,\ep) = \sum_{q,\om} Im G^R(q,\ep-\om) Im V^R(k-q,\om) \; 
\{\coth(\om/2T) \; + \; \tanh((\ep-\om)/2T) \}\;\;.
\ee
Taking 
\be
Im G^R(k,\ep) = -\pi \delta(s_{k,\ep}), \; \; s_{k,\ep} = \ep+\mu-\ep_k \;\; ,
\ee
we obtain
\be
Im \Sigma^R(k,\ep) = -\pi \sum_q Im V^R(k-q,s_{q,\ep}) \;\{ \coth(s_{q,\ep}/2T) \;
+ \; \tanh((\ep_q-\mu)/2T) \}\;\; .
\ee
Setting $Im G(k,\ep)$ equal to a delta function appears to be a reasonable
approximation for this purpose, since numerically it is 
{\em small} compared to the band energy - typically $Im G(k,\ep) < t/20 $
for $\{U_i\} \le t$, with a magnitude roughly {\em proportional} to $\{U_i\}$.
In fact, the difference of $Im \Sigma(k,\ep)$, as seen in our numerical
calculation, for small and large coupling constants is mostly {\em quantitative}
rather than qualitative ! 

	We assume that 
\be
Im V^R(q,x) = \sum_{n=0}^{\infty} \; \frac{V_q^{(2n+1)}(0) \; x^{2n+1}}{(2n+1)!}
\; ,
\ee
where $V_q^{(n)}(0)$ is the $n-$th derivative of $Im V(q,\om=0)$ with respect
to $\om$. This is true for an electronically mediated interaction,
with a polarization which is a regular function of $\om$ (see also below).

	First we consider the low $T$ limit. 
The sum over $q$ is dominated by the van-Hove singularities at the points 
$q_o$. Assuming that $\ep_{vH} = \ep_{q_o}$ is {\em close} to $\mu$, the
tanh has a vanishing contribution at the vicinity of $\ep_q \sim \mu$
(note that for $\ep_q < \mu$ and $\ep_q > \mu + \ep$ the contributions
of tanh and coth annihilate each other in the low $T$ limit). Hence
\be
Im \Sigma^R(k,\ep) \simeq -\pi \sum_{q \sim q_o} \sum_{n=0}^{\infty} 
\frac{ V_{k-q}^{(2n+1)}(0) \; (s_{q,\ep})^{2n+1}}{(2n+1)!} \; \;,
\ee
where we assumed that $\mu > \ep_{vH} = \ep_{q_o}$. 
For {\em sufficiently small} $V_{q}^{(n)}(0), \; \forall n>1$, we obtain
\be
a_k = \sum_{q \sim q_o} V_{k-q}^{(1)}(0) \; \gg \; \sum_{q \sim q_o} 
\sum_{n=1}^{\infty} \; \frac{ V_{k-q}^{(2n+1)}(0) \; (\ep + c)^{2n}}{(2n+1)!} 
\;\;,
\ee
where $c = \mu - \ep_{vH}$. This relation is valid for $\ep + c < \ep_c$,
where the latter is the characteristic energy beyond which the infinite sum
on the right becomes comparable to the $V_{q}^{(1)}$ term. 
Also, for energies beyond the bandwidth $8t$, $Im \chi_o$, and hence 
$Im V$ (see below), decay to zero. 
These considerations permit the identification
\be
\ep_1 = \mu - \ep_{vH} \;\;, \;\;
\ep_2 = \min\{\ep_c + \ep_{vH} - \mu, \; 8t + \ep_{vH} - \mu\} \;\; .
\ee
The assumption above for $a_k$ yields  
\be
Im \Sigma^R(k,\ep) \simeq -\pi a_k (\ep + c) \;\;, \ep_1 <  \ep  < \ep_2 \;\;.
\ee
Finally, we note that, if the Fermi surface approaches a van-Hove singularity
at $q_o$, 
$V_{k-q}^{(1)}(0)$ should become bigger, being proportional
to $1/(\vec \nabla \ep_{k_F} \vec k_F)$ (as implied by the standard Fermi liquid
result for the imaginary part of the susceptibility).

	We consider now the high temperature limit $T\gg \ep$. 
We see immediately that 
\be 
Im \Sigma^R(k,\ep) = -\pi \sum_q Im V^R(k-q,s_{q,\ep}) \{ 2 T/s_{q,\ep} +
O(s_{q,\ep}/2T) \} \;\; .
\ee
(Note that the term of order $T$ of this sum is
reminiscent of the left-hand side of the sum rule - 
c.f. Pines and Nozi\`eres\cite{pino} - 
$\lim_{q \rightarrow 0} \int_0^{\infty} d\om Im \chi_o(q,\om) |\varepsilon
(q,\om)|^2/\om = - N\pi/m c_s^2 $, with $N$ being the total particle number,
$c_s$ the speed of sound, $m$ the effective mass, and $\varepsilon(q,\om)$
the dielectric function.)
The sum is dominated by the van-Hove singularities at the points
$q_o$, thus yielding
\be
Im \Sigma^R(k,\ep) \simeq - 2 T \pi \sum_{q \sim q_o} V_{k-q}^{(1)}(0)  = 
- 2 \pi a_k T \;\;.
\ee
Here we made use of the condition above for $a_k$.

In all, we showed that the scattering rate $\tau^{-1}(k,\ep)$ of the 
quasiparticles obeys
\be
\tau^{-1}(\ep) \simeq a\ep \;,\; T\ll \ep \;\;, \;\tau^{-1}(T) \simeq bT \;,\; 
T\gg \ep
\;\;,
\ee 
with a ratio $b/a=2$.
We note that the $T$ and $\ep$ dependence of $\tau^{-1}_k$ 
is {\em independent}
of $k$ - thus {\bf leading necessarily to a linear in $T$ resistivity and a linear
in $1/\ep$ optical conductivity}, even with inclusion of vertex corrections
in the calculation.

	Numerically $Im V(q,\om)$ is seen to be linear in $\om$
for small $\om$, and then it has 2 or 3 distinct peaks, before decaying
to zero for $\om$ greater than the bandwidth $8t$. This behavior
closely follows $Im \chi_o(q,\om)$, as in fact
\be
Im V = Im \chi_o \;\; |\varepsilon(\{U_i\},Re \chi_o, Im \chi_o)|^2 \;\;,
\ee
with $\varepsilon$ being a rational function of its arguments.
The overall behavior of $Im V(q,\om)$ follows from any screened interaction
between the carriers. Hence the argument for the
linear in energy and temperature behavior of $\tau^{-1}(T,\ep)$
is equally generic. It relies 
on a large coefficient for the 
linear in energy term of $Im V$, and the presence of strong peaks in the
density of states - van-Hove singularities in 2-dimensions - 
{\em near} the chemical potential. What is more, in our numerical solution 
we observe that the energy $\ep_{vH}$ of the singularities is pushed 
by the interactions close to the chemical potential. The shape of the 
imaginary part of the self-energy $Im \Sigma(k,\ep)$ of the interacting system
being responsible for the modification of the density of states $N(\ep)$, 
through the relation $N(\ep)=-Tr \; Im G(k,\ep)/\pi$
($Im \Sigma(k,\ep)$ has a peak below $\mu$ and a dip above it, which 
account for the transfer of the spectral weight).
This feedback effect reinforces the role of the singularities, which 
are more effective in producing a linear scattering rate the closer they 
are to the Fermi surface. In fact this seems to be a plausible explanation
for the common characteristic of a good many cuprates whose van-Hove
singularities are located between 10-30 $meV$ below the Fermi surface
\cite{dlu} .

	It is interesting that the electron doped $Nd_{2-x}Ce_xCuO_{4+\delta}$
which has a van-Hove singularity much below the Fermi surface, i.e. at 
approximately $\mu$-350 meV, as shown by ARPES\cite{stanford},
has a usual Fermi liquid $\tau^{-1}(T) = const. \; T^2$ \cite{tsuei}.
This lends support to the picture described above.

	Let us note here that in all likelihood the present mechanism 
cannot explain the experimentally observed $T^2$ dependence of the Hall 
resistivity of the cuprates. A succesful
way to explain it has been found by Stojkovic and Pines \cite{stp}, 
using an electron interaction peaked at Q$=(\pm \pi,\pm \pi)$. 
Their argument can be slightly modified, so that it works for our form
of the electron potential $V$, but with a modified 
susceptibility peaked at Q, as we propose elsewhere \cite{eg}.

\vspace{1cm}

\centerline{\bf On the susceptibility of the cuprates}

\vspace{1cm}

	The Millis-Monien-Pines susceptibility\cite{pines}
\be
\chi_{MMP}(q,\om) = \frac{\xi^2 \; X_1}{1+\xi^2 (q-Q)^2 - i\om/\om_{SF}} \;\;,
\ee
has been used to fit the 
susceptibility of the cuprates for {\em low} $\om$ 
from both NMR and inelastic neutron scattering (INS) experiments.
The short range antiferromagnetic order, a remnant of the parent 
antiferromagnetic materials, with correlation length $\xi$, 
is responsible for the peak of the susceptibility for $q$ near $Q$.
Typically $\xi$ is of the order of the lattice constant, while
$\om_{SF} \approx 10-40 meV$.

	The origin of the small magnitude of $\om_{SF}$ has remained 
elusive thus far.
Chubukov, Sachdev and Sokol have interpreted it as a damped spin
wave mode \cite{css}. Spin waves are clearly observable in underdoped cuprates.
However to date, 
experimental evidence has it that spin waves are overdamped 
in the normal phase of the optimally doped and overdoped regimes. 

	An alternative explanation is a fermionic origin for $\om_{SF}$.
Suppose we write 
the total 
susceptibility as
\be
\chi(q,\om) = \chi_{AF}(q,\om) + \chi_o(q,\om)\;\;, \;\;q \rightarrow Q ,\;\; 
\ee
\be
\chi_{AF}(q,\om) = \frac{\xi^2 \; \chi_1}{1+\xi^2 (q-Q)^2 - f(\om)} \;\;, \;\;
\chi_o(q,\om) = \frac{\chi_o}{1-i \om/\om_o} \;\;.
\ee
$\chi_o$ is the fermionic susceptibility, while $\chi_{AF}$
has an antiferromagnetic origin - being due, e.g., to the localized Cu spins.
Note that a similar proposal for the susceptibility has been put forward
by Onufrieva and Rossat-Mignod\cite{onu}. 
If $|f(\om)|\ll \om/\om_o$ we recover essentially 
$\chi_{MMP}(q,\om)$ - which is itself
an approximate form of the true susceptibility - with $\om_{SF}=\om_o$.

	Hence the origin of small $\om_o(\vec q)=\vec \nabla \ep_{k_F}
\vec q$ 
is the proximity of the
Fermi surface to a van-Hove singularity - c.f. the discussion in the
context of the scattering rate above. From the numerical solution
of our system, we easily obtain values of $\om_o$ comparable to 
the experimentally relevant ones, when the chemical potential is
near $\ep_{vH}$. 
Actually this is the case with a good number of cuprates.
In ref. \cite{dlu} there is a compilation of several cuprates,
the van-Hove singularities of which are located between 10 - 30 $meV$ 
below the Fermi level.
Also, Blumberg, Stojkovic and Klein (BSK)\cite{bsk} suggest 
that this characteristic may be true irrespective of the doping,
as long as the latter is appropriate for superconductivity.
This is based on ARPES experiments on the bilayer $YBa_2Cu_2O_{7-\delta}$. 
Unfortunately, 
it has not proved possible yet to perform measurements
on many compounds, especially the monolayers $La_{1-x}Sr_xCuO_4$
and $Tl_2Ba_2CuO_{6+\delta}$.
The point here is the following. By fitting the ARPES data, BSK
show that one of the two effective bands - the anti-bonding one - formed 
by hybridization of the two layers by interlayer coupling has a chemical
potential only some 20 - 50 $meV$ above the van-Hove singularity at
$(0,\pi)$, irrespective of the doping regime. It is then clear that these
carriers, with a large density of states, give rise to a {\em small} $\om_o$ 
as discussed above. Hence it is very 
interesting to know 
how universal this 
characteristic of the cuprates is,
as it may explain naturally the magnitude of $\om_o$.
Furthermore, it would be interesting to determine experimentally,
eg. by NMR, the value of $\om_o$ for $Nd_{2-x}Ce_xCuO_{4+\delta}$.
In that case, $\om_o$ should be enhanced as a result of the van-Hove 
singularities being far away from the Fermi surface.

	Finally, we have checked that within a weakly-disordered
model for the planes, we do {\em not} obtain a small $\om_o$ \cite{eg}, as might
have been expected. The rationale behind this model is
the disorder inherent in the cuprates.
The dopants are randomly positioned in the crystal structure, thereby 
creating an effective disorder potential for the carriers in the planes.

\vspace{1cm}

\centerline{\bf Summary}

\vspace{1cm}

	In summary, we have shown that a fermionic model, with strong 
van-Hove singularities located close to the chemical potential $\mu$,
is able to account in a natural and consistent manner for a number of
characteristics of the optimally doped and overdoped cuprates.
The list includes the linear in $T$ resistivity, the linear in $1/\ep$
optical conductivity, the pinning of the van Hove-singularities close
to $\mu$, and the low energy dependence of the susceptibility.
Existing experimental evidence is quite favorable for the model,
while some new experiments are proposed in light of the model.
We will report elsewhere \cite{eg} on the implications of our approach 
for the superconducting transition.

\vspace{1cm}

	The author has enjoyed useful discussions with 
Yia-Chung Chang, Gordon Baym,
Girsh Blumberg, Antonio Castro Neto, Tony Leggett, Qimiao Si and 
Branko Stojkovic. This work was supported by the Research Board of 
the University of Illinois, the Office of Naval Research and NSF.

\end{document}